\documentclass[twocolumn,nofootinbib,amsmath,amssymb,aps,prd,balancelastpage]{revtex4-1}

\usepackage{amsmath,amssymb}
\usepackage{amsfonts,amssymb,mathrsfs}

\usepackage[bookmarks=false,pdfstartview=FitH]{hyperref}
\usepackage[all]{hypcap}

\def\be{\begin{equation}}
\def\ee{\end{equation}}

\def\f{\frac}

\def\dd{{\rm d}}
\def\al{\alpha}

\def\ep{\epsilon}

\def\om{\omega}

\def\mR{\mathcal{R}}
\def\mH{\mathcal{H}}

\def\mP{\mathcal{P}}

\def\mN{\mathcal{N}}

\begin{document}

\pagestyle{plain}

\title{Running of the scalar spectral index in bouncing cosmologies}

\author{Jean-Luc Lehners} \email{jean-luc.lehners@aei.mpg.de}
\affiliation{Max Planck Institute for Gravitational Physics (Albert Einstein Institute),\\
Am M\"uhlenberg 1, 14476 Golm, Germany, EU}

\author{Edward Wilson-Ewing} \email{wilson-ewing@aei.mpg.de}
\affiliation{Max Planck Institute for Gravitational Physics (Albert Einstein Institute),\\
Am M\"uhlenberg 1, 14476 Golm, Germany, EU}

\begin{abstract}
We calculate the running of the scalar index in the ekpyrotic and matter bounce cosmological scenarios, and find that it is typically negative for ekpyrotic models, while it is typically positive for realizations of the matter bounce where multiple fields are present.  This can be compared to inflation, where the observationally preferred models typically predict a negative running.  The magnitude of the running is expected to be between $10^{-4}$ and up to $10^{-2},$ leading in some cases to interesting expectations for near-future observations.
\end{abstract}

\maketitle

\section{Introduction}
\label{s.intro}

Observations from Planck \cite{Ade:2015xua,Ade:2015lrj} indicate that primordial perturbations were adiabatic, near-Gaussian, almost scale-invariant with a slight red tilt, and that tensor perturbations were small.  These observations have already greatly constrained the dynamics of the early universe.

Our goal is to consider the running of the scalar spectral index as a further observational tool that can be used to differentiate between various cosmological scenarios, and in particular between inflation and two alternatives, the ekpyrotic universe and the matter bounce scenario.  Even though the bounds obtained from observations already do provide interesting constraints on these models, it will be helpful to have other observable quantities at our disposal to distinguish between these models, especially since non-Gaussianities are already known to be small and that tensor modes have remained elusive.

The primordial scalar perturbations are usefully characterised via their power spectrum $\mP_\mR(k),$ which describes the root mean square amplitude of the scalar curvature perturbations as a function of their wavenumber $k.$ The spectral index of the perturbations is defined via
\be
n_s - 1 \equiv \f{\dd \ln \mP_\mR}{\dd \ln k}\bigg|_{\text{at } k=a|H|} = - \f{\dd \ln \mP_\mR}{\dd \mN},
\ee
and it tells us how the amplitude of the perturbations is changing with scale. Here $a$ denotes the scale factor of the universe, $H=\dot{a}/a$ the Hubble rate and we have defined ${\cal N}$ as the number of e-folds left before the end of inflation, ekpyrosis or matter contraction,
\begin{equation}
\mN_{tot} - \mN \equiv \ln (a|H|) \,.
\end{equation}
For inflation, given that the Hubble rate $H$ is nearly constant, one often defines the number of e-folds in terms of the variation of the scale factor alone. However, in bouncing cosmologies such an approximation is inappropriate, as the growth of $|H|$ is crucial. Note that in order for the flatness and horizon problems to be resolved, it is precisely the quantity $a|H|$ that must grow by a factor $e^{60}$ or more before the hot big bang expansion phase. Observations now indicate a clear deviation from a purely scale-invariant spectrum $n_s=1,$ with Planck indicating a value \cite{Ade:2015xua,Ade:2015lrj}
\be
n_s = 0.968 \pm 0.006, \quad (68\% \text{ CL}),
\ee
from the combined data from temperature fluctuations and lensing. Given that the spectrum has a slightly different amplitude at different scales, one may wonder whether the deviation from scale invariance also changes with scale. This is captured by the running $\alpha_s$ of the spectral index, defined via
\be
\al_s \equiv \f{\dd n_s}{\dd \ln k}\bigg|_{\text{at } k=a|H|}\, = -\f{\dd n_s}{\dd \mN},
\ee
Given that in all the models we are considering here the longest-wavelength modes get produced first, $\alpha_s$ has the same sign as the time derivative of the spectral index, $\dd n_s/\dd t.$ Such a running has not been observed yet, but Planck provides the following bound (again from the combined data from temperature fluctuations and lensing \cite{Ade:2015xua,Ade:2015lrj})
\be \label{PlanckRunning}
\alpha_s = -0.003 \pm 0.007, \quad (68\% \text{ CL}).
\ee
We are interested in how this measured bound compares with various theoretical models, and how future observations may allow one to discriminate between models. We will begin by reviewing the predictions for inflationary models, and then contrast these with bouncing cosmologies.  Note that for the purposes of the present paper, we will consider the na\"ive predictions of these cosmological models and assume (perhaps optimistically given that these issues are not fully understood yet) that they are not significantly modified by eternal inflation or the properties of the bounce.

\section{Review of the Running of the Scalar Index in Inflation}
\label{s.inf}

The power spectrum of (single field) inflationary models is given, up to a numerical factor, by
\be
\mP_\mR(k) = \f{H^2}{\ep},
\ee
where $H$ is the Hubble rate during inflation and $\ep$ is the slow-roll parameter
\be \label{eps}
\ep \equiv -\f{\dd \ln |H|}{\dd \ln a}.
\ee
From here on, we will ignore corrections at sub-leading order in $\ep.$  Then, the scalar index $n_s$ is given by (see e.g. \cite{Wang:1997cw})
\begin{eqnarray}
n_s - 1 =  - \f{\dd \ln \mP_\mR}{\dd \mN} &=& - \frac{2 \, \ep}{1-\ep} + \f{\ep_{,\mN}}{\ep} \nonumber \\ &=& -2 \, \ep + \f{\ep_{,\mN}}{\ep} + {\cal O}(\ep^2),
\end{eqnarray}
while the running is
\begin{eqnarray} \label{inf}
\alpha_s = - \f{\dd n_s}{\dd \mN}
= 2 \, \ep_{,\mN} - \f{\ep_{,\mN\mN}}{\ep} + \left( \f{\ep_{,\mN}}{\ep} \right)^2.
\end{eqnarray}

A very useful way of parameterising inflationary models is obtained by combining the demand for a graceful exit ($\ep=1$ when $\mN=0$) with the assumption of a scaling symmetry \cite{Mukhanov:2013tua,Ijjas:2013sua},
\be
\ep=\frac{1}{(\mN + 1)^p}\,.
\ee
This parameterisation captures the qualitative behaviour of both chaotic ($p = 1$) and plateau ($p = 2$) inflationary models, and it may be used to derive the predictions of inflationary models with ``minimal'' tuning. With this parameterisation, the spectral index and its running come out as
\begin{eqnarray}
n_s -1 &=& -\frac{2}{(\mN+1)^p} - \frac{p}{\mN + 1}\,, \\
\alpha_s &=&  -\frac{2p}{(\mN+1)^{p+1}} - \frac{p}{(\mN + 1)^2}\,.
\end{eqnarray}
For chaotic models ($p\approx 1$), this leads to the typical values $n_s \approx 0.95$ and $\alpha_s\approx -8\times 10^{-4},$ assuming $\mN=60$ for the modes of interest. For the plateau models favoured by current observations ($p\approx 2$), we find the typical values $n_s\approx 0.97$ and $\alpha_s \approx -5\times 10^{-4}.$ In both cases (in fact in {\it all} cases captured by the above parameterisation) the running is negative in sign and of magnitude ${\cal O}(1/\mN^2)$ (note that in order to match the observed tilt in $n_s$, $p$ must lie between approximately 1 and 3).

These results are simply presenting in a different fashion the well-known statement that, while inflation typically predicts a negative running \cite{Kosowsky:1995aa}, a large negative running can be problematic for inflation \cite{Easther:2006tv}.  That being said, it is also known that if the inflaton potential contains modulations, this can increase the amplitude of the running (and even potentially give a positive running) \cite{Kobayashi:2010pz}.  The drawback here is that this necessarily requires adding extra parameters to the theory, making it less predictive.  Furthermore, the presence of sufficiently large modulations may make a graceful exit from inflation more difficult.

\section{The Ekpyrotic Universe}
\label{s.ek}

Instead of having an expansion phase with substantial negative pressure  (inflation), the cosmological flatness problem may alternatively be solved by having a contracting phase with large positive pressure (ekpyrosis). The high pressure suppresses both the homogenous curvature in the universe as well as small anisotropies, thus rendering the universe increasingly flat as it contracts \cite{Erickson:2003zm}. A successful model then requires linking the ekpyrotic phase via a bounce to the current expanding phase of the universe --- for implementations of smooth bounces see e.g.\ \cite{Buchbinder:2007ad, Creminelli:2007aq, Easson:2011zy, Cai:2012va, Qiu:2013eoa, Koehn:2013upa, Wilson-Ewing:2013bla} and for a proof that the perturbations of interest pass through such bounces unaltered see \cite{Wilson-Ewing:2013bla, Battarra:2014tga}.

In single-field ekpyrotic models, scalar perturbations are not amplified and thus the universe remains perfectly smooth, too smooth in fact! However, two-field models can produce scalar perturbations by amplifying isocurvature/entropy modes first, and subsequently converting them into adiabatic curvature perturbations. Two types of such models have been considered to date, and we will analyse them in turn. For the first type, one introduces a potential which is unstable in the field space direction perpendicular to the ekpyrotic direction (which is the entropic direction) \cite{Notari:2002yc,Finelli:2002we,Lehners:2007ac}. This instability is responsible for amplifying the entropic modes, but it also implies that the ekpyrotic phase now requires special initial conditions (although in a cyclic context, this may be a feature rather than a problem \cite{Lehners:2008qe}). As detailed in \cite{Lehners:2007ac}, the spectral index of the perturbations comes out as
\be
n_s -1 = \frac{2}{\epsilon} - \frac{\epsilon_{,\mN}}{\epsilon}\,. 
\ee 
Here $\epsilon = - \dd \ln H/\dd \ln a$ is defined in the same way as in inflation, with the difference that in the ekpyrotic case $\ep$ is required to be larger than $3$ (i.e., the potential is required to be steep). The calculation of the running then follows in a straightforward fashion, with the result that
\be
\alpha_s = \frac{2 \, \ep_{,\mN}}{\ep^2} + \frac{\ep_{,\mN\mN}}{\ep} - \left(\frac{\ep_{,\mN}}{\ep}\right)^2\,.
\ee
For ekpyrosis, we can apply the same requirements of graceful exit ($\ep=3$ when $\mN=0$) and scaling symmetry to obtain the parameterisation \cite{Ijjas:2013sua}
\be
\ep = 3(\mN+1)^p\,.
\ee
This leads to the following expressions for the spectral index and the running:
\begin{eqnarray}
n_s -1 &=& \frac{2}{3(\mN+1)^p} - \frac{p}{\mN + 1}\,, \\
\alpha_s &=&  \frac{2p}{3(\mN+1)^{p+1}} - \frac{p}{(\mN + 1)^2}\,.
\end{eqnarray}
Models fitting the observations well have $p\approx 2,$ for which $n_s \approx 0.97.$ For these models, we may expect the running to be given by $\alpha_s \approx -5\times 10^{-4}.$ These predictions are in fact essentially identical to those of plateau models of inflation (also obtained for $p\approx 2$). It is interesting to observe that in the present class of models all cases leading to a red tilt ($p\gtrapprox 1$) also lead to a negative running of the spectrum. We should also remark that models of this type can produce significant non-Gaussian corrections \cite{Lehners:2007wc,Lehners:2009ja}, which are in agreement with Planck bounds if the unstable potential is symmetric with respect to the ekpyrotic field direction \cite{Lehners:2013cka}. 

The second type of ekpyrotic model we will consider are models where the second scalar field does not acquire a potential (and remains stable), but where this scalar is non-minimally coupled to the ekpyrotic field $\phi.$ The matter Lagrangians we have in mind are of the form \cite{Ijjas:2014fja}
\be
{\cal L} = \sqrt{-g}\left[- \frac{1}{2}(\partial\phi)^2 - \frac{1}{2} \Omega(\phi)^2 (\partial \chi)^2 - V(\phi) \right]\,,
\ee 
where the second scalar is denoted $\chi$ and it is kinetically coupled to the ekpyrotic field via a function $\Omega(\phi).$ As shown in \cite{Ijjas:2014fja}, for each ekpyrotic potential $V(\phi)$ there exists a coupling function $\Omega(\phi)$ such that the perturbations in $\chi$ obtain a specified spectrum, e.g.\ scale-invariant, but large departures from scale invariance are also possible. In fact, if we allow for arbitrary tuning, we may always find a function $\Omega$ that yields a given spectral index and a given running. This is similar to the situation in inflation, where, allowing arbitrary tuning of the potential, virtually any such combination may also be obtained. However, we are interested in models that are simple and that require as little fine-tuning as possible. For this reason, we will consider the case where the potential can be approximated by the standard ekpyrotic form $V(\phi) = - V_0 e^{-c\phi},$ but where we allow $c$ to vary slowly over time so as to allow the ekpyrotic phase to have a graceful exit. Then for a coupling function $\Omega^2 = e^{-b\phi}$ with $b$ being a constant, the spectral index comes out as \cite{Fertig:2013kwa}
\begin{equation} \label{nsnonminimal}
n_s-1 = -2 \left( \frac{b}{c} - 1\right) - \frac{7 \, \epsilon_{,\mN}}{3 \epsilon}\,,
\end{equation} 
where we have assumed that $\epsilon = c^2/2$ is large and we have dropped terms sub-leading in $1/\epsilon.$ The situation we have in mind is where $b$ and $c$ are equal at the beginning of the ekpyrotic phase, and where $c$ then gradually decreases according to $\epsilon = c^2/2 = 3(\mN+1)^p.$ In this case, part of the observed red tilt can already be generated by the first term on the right hand side of (\ref{nsnonminimal}), implying that we would need to take $p$ small, certainly $p<1.$ For this model, the running is given by
\begin{eqnarray}
\alpha_s &=& - \frac{b \, \ep_{\mN}}{c \, \ep} + \frac{7\, \ep_{,\mN\mN}}{3 \ep} - \frac{7}{3}\left( \frac{\ep_{,\mN}}{\ep}\right)^2\,, \\ &=& - \frac{b p}{c (\mN+1)} - \frac{7 p}{3 (\mN+1)^2}\,.
\end{eqnarray} 
Again, the running is expected to be negative, and in fact it may be quite large in this class of models. For the extreme case where the entire red tilt is generated by the variation of $\epsilon$, we have $p\approx 1$ and thus a very large running of $\alpha_s\approx -10^{-2},$ which is at a level that can be seen by near-future experiments. For the more general case where some of the red tilt is generated by the difference between $b$ and $c$, we would expect $p$ to be smaller and $\alpha_s$ correspondingly also, but still at a significant level of $\alpha_s \approx - 10^{-3}$ for values of $p$ within the range of validity of the approximations used here (since the spectral index was derived using the approximation that $\epsilon$ is large, we cannot take $p$ smaller than approximately $p=1/10$). Hence this type of model is special in that it predicts a much more significant running than the inflationary and ekpyrotic models discussed up to now, and this is something to look out for in upcoming observations.  

Recently, new cosmological models have been proposed (anamorphosis \cite{Ijjas:2015zma} and conflation \cite{Fertig:2015dva}) that combine in various ways elements of both inflation and ekpyrosis. Based on the arguments above we expect the running to also be negative for these models, while the amplitude of $\al_s$ will depend on the specific realization.

\section{The Matter Bounce Scenario}
\label{s.mb}

As discovered by Wands \cite{Wands:1998yp}, a simple matter dominated contracting phase can also lead to scale-invariant scalar and tensor perturbations. This observation allows the construction of cosmological models in which the perturbations are generated in the presence of ordinary pressure-free matter \cite{Finelli:2001sr}.  The horizon problem is automatically solved in matter bounce models since the big-bang singularity is resolved, and the flatness problem is alleviated: since the initial conditions are set far in the pre-bounce phase, it is possible to postulate a similar amplitude of spatial curvature prior to the bounce point as is observed today \cite{Brandenberger:2012zb}.  Note however that in a contracting FLRW space-time, a matter phase is not an attractor solution, and in particular the matter bounce is sensitive to the growth of anisotropies which in the absence of a fine-tuning of the initial conditions will come to dominate the high-curvature dynamics for generic solutions \cite{Xue:2010ux, Xue:2011nw} (although this last problem can be mitigated by adding an ekpyrotic phase after the matter-dominated era \cite{Cai:2013vm}).

In addition, and just as for ekpyrosis, the model must be completed by a description of a bounce. Realizations of the matter bounce scenario typically fall into two categories depending on whether the bounce is caused by exotic matter fields that violate energy conditions \cite{Cai:2008qw, Lin:2010pf, Cai:2012va} or by modifications of the Einstein equations that are typically motivated by quantum gravity \cite{Brandenberger:2009yt, Biswas:2005qr, WilsonEwing:2012pu, Brandenberger:2013zea, Peter:2006hx, Odintsov:2014gea, Cai:2014zga, Bamba:2014mya, Cai:2014jla}.  Note that the bounce can play an important role: for instance, bounces in loop quantum cosmology \cite{WilsonEwing:2012pu, Cai:2014jla} tend to suppress tensor fluctuations relative to scalar ones, which is a beneficial feature here as a contracting matter phase by itself tends to produce tensor fluctuations with an amplitude above present observational bounds \cite{deHaro:2014kxa}.

Here, we will concentrate on the scalar fluctuations. We will briefly review (and extend by allowing for a dynamical $\ep$) the calculation of their spectral index \cite{Peter:2006hx, WilsonEwing:2012pu, Elizalde:2014uba, deHaro:2015wda} in order to calculate the expected running. The mode functions $v\equiv z {\cal R}$ of the gauge-invariant curvature perturbations ${\cal R}$ obey the Fourier space evolution equation (for sound speed $c_s = 1$)
\begin{equation}
v^{\prime\prime} + \left(k^2 - \frac{z^{\prime\prime}}{z}\right) v = 0\,,
\end{equation}
where a prime denotes a derivative with respect to the conformal time $\tau,$ $k$ is the comoving wavenumber and $z=a(2\epsilon)^{1/2}.$  It is easy to verify that, neglecting terms with two $N$ derivatives on $\ep$ (note that $\ep''/2 \ep$ contributes a term linear in $\ep_{,N}$),
\begin{equation}
\frac{z^{\prime\prime}}{z} ={\cal H}^2 \left( 2-\ep + \frac{3 \, \ep_{,N}}{2 \ep} -\frac{\ep_{,N}}{2} \right) \,,
\end{equation}
where $\mH = a'/a$, $\dd N=\dd\ln(a)$ (or equivalently $\dd\mN=(\ep - 1) \dd N$), and we have used the relation $a''/a = \mH^2 (2-\ep)$ which follows from \eqref{eps}.

An expression for ${\cal H}$ can be obtained from the definition ${\cal H}^\prime = {\cal H}^2 (1-\ep)$, which after successive integrations by parts (and neglecting terms of the order $\ep_{,NN}$ and $(\ep_{,N})^2$) gives \cite{Lehners:2007ac}
\begin{equation}
\frac{1}{{\cal H}} = (\ep-1) \tau \left[1-\frac{\ep_{,N}}{(\ep-1)^2} \right]\,.
\end{equation}
Putting everything together (in the approximation that second order $N$ derivatives are negligible),
\begin{equation}
\frac{z^{\prime\prime}}{z} =\frac{\left( 2-\ep + \frac{3 \ep_{,N}}{2 \ep} -\frac{\ep_{,N}}{2} + \frac{4 \ep_{,N}}{(\ep-1)^2} - \frac{2 \ep\ep_{,N}}{(\ep-1)^2}\right)}{(\ep-1)^2 \, \tau^2}  \,.
\end{equation}
Requiring the fluctuations to initially be in their vacuum state in the far past, the solution for the mode functions is given (up to an irrelevant phase) by
\be
v= \left(-\frac{\pi \tau}{4} \right)^{1/2}  H_\nu^{(1)}(-k\tau)\,,
\ee
where $H_\nu^{(1)}$ is a Hankel function with index 
\begin{equation}
\nu = \frac{1}{2(\ep-1)}\left( 3-\ep+\frac{\ep_{,N}}{\ep}+\frac{4(\ep-2)\ep_{,N}}{(\ep-3)(\ep-1)^2} \right)\,.
\end{equation}
Rewriting this in terms of the effective equation of state of the matter fields $\om$ (related to $\ep$ by $\om = \f{2}{3} \ep - 1$) and $\mN$ derivatives, we finally obtain
\begin{equation} \label{mattertilt}
n_s - 1 = 12 \, \om - 9 \, \om_{,\mN} \,,
\end{equation} 
and an immediate consequence is that the running can be approximated by
\begin{equation}
\alpha_s = -12 \, \om_{,\mN} \,.
\end{equation}
There is no simple way to estimate the quantitative behaviour of the equation of state, unlike for the inflationary and ekpyrotic models above. However, given that matter components with larger and larger equation of state successively come to dominate in a contracting universe, we would expect the effective equation of state to grow during the matter phase, i.e. as $\mN$ shrinks. Thus it is natural to expect $\om_{,\mN} < 0$ which would give a positive running, in contrast with inflation and ekpyrosis. 

A positive running is thus a typical signature of a matter bounce model where multiple fields are present.  However, if there is only one field contributing to the dynamics, it is possible to obtain a negative running.  One such example is given by models where it is a scalar field with an exponential potential that is acting as the pressure-less matter field \cite{deHaro:2015wda}; the negative running arises in this case because, as mentioned above, the pressure-less solution is not an attractor in a contracting space-time and in some cases the equation of state will shrink before growing again to reach the attractor which is kinetic-dominated \cite{Heard:2002dr}.  Nonetheless, in this paper we are interested in ``typical'' realizations of the matter bounce and so long as there are a number of matter fields present in the space-time, one expects that the equation of state will generically increase in a contracting FLRW space-time.

How large do we expect the running to be? The amplitude of the running in the matter bounce varies significantly from one model to another, but in any  case the observational bound on $\al_s$ gives the constraint
\be
|\om_{,\mN}| < 10^{-3}, \quad (68\% \text{ CL}).
\ee
One important consequence is that the main contribution to the red tilt of the scalar index, given in \eqref{mattertilt}, must come from a negative $\om$. Thus the spectral running provides both an interesting signature of the matter bounce, as well as a strong constraint for specific realizations of the matter bounce.

\section{Discussion}
\label{s.disc}

We have analyzed the predictions for the running of the scalar spectral index in ekpyrotic and matter bounce models, and contrasted these predictions with those obtained for simple inflationary models.  The arguments given here are for ``typical'' realizations of the various cosmological scenarios. In that case we find that in all models the magnitude of spectral running is expected to be of the order of a few times $1/\mN^2$ or about $10^{-3}$ for the modes of cosmological interest, and for non-minimally coupled ekpyrotic models and for the matter bounce the magnitude of the running may even be as large as $10^{-2}$. For inflationary and ekpyrotic models, the running is expected to be negative, while for the matter bounce it is typically positive. Deviations from these predictions will require the models to be more complicated than the simple parameterisations we have considered here (but might point to interesting missing ingredients), while an experimental confirmation of these predictions would provide a satisfying consistency test.

\acknowledgments

We would like to thank Yi-Fu Cai for helpful discussions and Robert Brandenberger for helpful comments on an earlier draft of this paper.
JLL gratefully acknowledges the support of the European Research Council via the Starting Grant Nr. 256994 ``StringCosmOS''.
EWE is supported by a grant from the John Templeton Foundation.

\raggedright


\begin{thebibliography}{10}

\bibitem{Ade:2015xua}
{\bf Planck}, P.~Ade {\em et al.}, ``{Planck 2015 results. XIII. Cosmological
  parameters},''
\href{http://arXiv.org/abs/1502.01589}{{\tt arXiv:1502.01589}}.

\bibitem{Ade:2015lrj}
{\bf Planck}, P.~Ade {\em et al.}, ``{Planck 2015 results. XX. Constraints on
  inflation},''
\href{http://arXiv.org/abs/1502.02114}{{\tt arXiv:1502.02114}}.

\bibitem{Wang:1997cw}
L.-M. Wang, V.~F. Mukhanov, and P.~J. Steinhardt, ``{On the problem of
  predicting inflationary perturbations},'' Phys.\ Lett.\ {\bf B414} (1997)
  18--27,
\href{http://arXiv.org/abs/astro-ph/9709032}{{\tt arXiv:astro-ph/9709032}}.

\bibitem{Mukhanov:2013tua}
V.~Mukhanov, ``{Quantum Cosmological Perturbations: Predictions and
  Observations},'' Eur.\ Phys.\ J.\ {\bf C73} (2013) 2486,
\href{http://arXiv.org/abs/1303.3925}{{\tt arXiv:1303.3925}}.

\bibitem{Ijjas:2013sua}
A.~Ijjas, P.~J. Steinhardt, and A.~Loeb, ``{Scale-free primordial cosmology},''
  Phys.\ Rev.\ {\bf D89} (2014) 023525,
\href{http://arXiv.org/abs/1309.4480}{{\tt arXiv:1309.4480}}.

\bibitem{Kosowsky:1995aa}
A.~Kosowsky and M.~S. Turner, ``{CBR anisotropy and the running of the scalar
  spectral index},'' Phys.\ Rev.\ {\bf D52} (1995) 1739--1743,
\href{http://arXiv.org/abs/astro-ph/9504071}{{\tt arXiv:astro-ph/9504071}}.

\bibitem{Easther:2006tv}
R.~Easther and H.~Peiris, ``{Implications of a Running Spectral Index for Slow
  Roll Inflation},'' JCAP {\bf 0609} (2006) 010,
\href{http://arXiv.org/abs/astro-ph/0604214}{{\tt arXiv:astro-ph/0604214}}.

\bibitem{Kobayashi:2010pz}
T.~Kobayashi and F.~Takahashi, ``{Running Spectral Index from Inflation with
  Modulations},'' JCAP {\bf 1101} (2011) 026,
\href{http://arXiv.org/abs/1011.3988}{{\tt arXiv:1011.3988}}.

\bibitem{Erickson:2003zm}
J.~K. Erickson, D.~H. Wesley, P.~J. Steinhardt, and N.~Turok, ``{Kasner and
  mixmaster behavior in universes with equation of state w $\ge$ 1},'' Phys.\
  Rev.\ {\bf D69} (2004) 063514,
\href{http://arXiv.org/abs/hep-th/0312009}{{\tt arXiv:hep-th/0312009}}.

\bibitem{Buchbinder:2007ad}
E.~I. Buchbinder, J.~Khoury, and B.~A. Ovrut, ``{New Ekpyrotic cosmology},''
  Phys.\ Rev.\ {\bf D76} (2007) 123503,
\href{http://arXiv.org/abs/hep-th/0702154}{{\tt arXiv:hep-th/0702154}}.

\bibitem{Creminelli:2007aq}
P.~Creminelli and L.~Senatore, ``{A Smooth bouncing cosmology with scale
  invariant spectrum},'' JCAP {\bf 0711} (2007) 010,
\href{http://arXiv.org/abs/hep-th/0702165}{{\tt arXiv:hep-th/0702165}}.

\bibitem{Easson:2011zy}
D.~A. Easson, I.~Sawicki, and A.~Vikman, ``{G-Bounce},'' JCAP {\bf 1111} (2011)
  021,
\href{http://arXiv.org/abs/1109.1047}{{\tt arXiv:1109.1047}}.

\bibitem{Cai:2012va}
Y.-F. Cai, D.~A. Easson, and R.~Brandenberger, ``{Towards a Nonsingular
  Bouncing Cosmology},'' JCAP {\bf 1208} (2012) 020,
\href{http://arXiv.org/abs/1206.2382}{{\tt arXiv:1206.2382}}.

\bibitem{Qiu:2013eoa}
T.~Qiu, X.~Gao, and E.~N. Saridakis, ``{Towards anisotropy-free and nonsingular
  bounce cosmology with scale-invariant perturbations},'' Phys.\ Rev.\ {\bf
  D88} (2013) 043525,
\href{http://arXiv.org/abs/1303.2372}{{\tt arXiv:1303.2372}}.

\bibitem{Koehn:2013upa}
M.~Koehn, J.-L. Lehners, and B.~A. Ovrut, ``{Cosmological super-bounce},''
  Phys.\ Rev.\ {\bf D90} (2014) 025005,
\href{http://arXiv.org/abs/1310.7577}{{\tt arXiv:1310.7577}}.

\bibitem{Wilson-Ewing:2013bla}
E.~Wilson-Ewing, ``{Ekpyrotic loop quantum cosmology},'' JCAP {\bf 1308} (2013)
  015,
\href{http://arXiv.org/abs/1306.6582}{{\tt arXiv:1306.6582}}.

\bibitem{Battarra:2014tga}
L.~Battarra, M.~Koehn, J.-L. Lehners, and B.~A. Ovrut, ``{Cosmological
  Perturbations Through a Non-Singular Ghost-Condensate/Galileon Bounce},''
  JCAP {\bf 1407} (2014) 007,
\href{http://arXiv.org/abs/1404.5067}{{\tt arXiv:1404.5067}}.

\bibitem{Notari:2002yc}
A.~Notari and A.~Riotto, ``{Isocurvature perturbations in the ekpyrotic
  universe},'' Nucl.\ Phys.\ {\bf B644} (2002) 371--382,
\href{http://arXiv.org/abs/hep-th/0205019}{{\tt arXiv:hep-th/0205019}}.

\bibitem{Finelli:2002we}
F.~Finelli, ``{Assisted contraction},'' Phys.\ Lett.\ {\bf B545} (2002) 1--7,
\href{http://arXiv.org/abs/hep-th/0206112}{{\tt arXiv:hep-th/0206112}}.

\bibitem{Lehners:2007ac}
J.-L. Lehners, P.~McFadden, N.~Turok, and P.~J. Steinhardt, ``{Generating
  ekpyrotic curvature perturbations before the big bang},'' Phys.\ Rev.\ {\bf
  D76} (2007) 103501,
\href{http://arXiv.org/abs/hep-th/0702153}{{\tt arXiv:hep-th/0702153}}.

\bibitem{Lehners:2008qe}
J.-L. Lehners and P.~J. Steinhardt, ``{Dark Energy and the Return of the
  Phoenix Universe},'' Phys.\ Rev.\ {\bf D79} (2009) 063503,
\href{http://arXiv.org/abs/0812.3388}{{\tt arXiv:0812.3388}}.

\bibitem{Lehners:2007wc}
J.-L. Lehners and P.~J. Steinhardt, ``{Non-Gaussian density fluctuations from
  entropically generated curvature perturbations in Ekpyrotic models},'' Phys.\
  Rev.\ {\bf D77} (2008) 063533,
\href{http://arXiv.org/abs/0712.3779}{{\tt arXiv:0712.3779}}.

\bibitem{Lehners:2009ja}
J.-L. Lehners and S.~Renaux-Petel, ``{Multifield Cosmological Perturbations at
  Third Order and the Ekpyrotic Trispectrum},'' Phys.\ Rev.\ {\bf D80} (2009)
  063503,
\href{http://arXiv.org/abs/0906.0530}{{\tt arXiv:0906.0530}}.

\bibitem{Lehners:2013cka}
J.-L. Lehners and P.~J. Steinhardt, ``{Planck 2013 results support the cyclic
  universe},'' Phys.\ Rev.\ {\bf D87} (2013) 123533,
\href{http://arXiv.org/abs/1304.3122}{{\tt arXiv:1304.3122}}.

\bibitem{Ijjas:2014fja}
A.~Ijjas, J.-L. Lehners, and P.~J. Steinhardt, ``{General mechanism for
  producing scale-invariant perturbations and small non-Gaussianity in
  ekpyrotic models},'' Phys.\ Rev.\ {\bf D89} (2014) 123520,
\href{http://arXiv.org/abs/1404.1265}{{\tt arXiv:1404.1265}}.

\bibitem{Fertig:2013kwa}
A.~Fertig, J.-L. Lehners, and E.~Mallwitz, ``{Ekpyrotic Perturbations With
  Small Non-Gaussian Corrections},'' Phys.\ Rev.\ {\bf D89} (2014) 103537,
\href{http://arXiv.org/abs/1310.8133}{{\tt arXiv:1310.8133}}.

\bibitem{Ijjas:2015zma}
A.~Ijjas and P.~J. Steinhardt, ``{The anamorphic universe},''
\href{http://arXiv.org/abs/1507.03875}{{\tt arXiv:1507.03875}}.

\bibitem{Fertig:2015dva}
A.~Fertig, J.-L. Lehners, and E.~Mallwitz, ``{Conflation: a new type of
  accelerated expansion},''
\href{http://arXiv.org/abs/1507.04742}{{\tt arXiv:1507.04742}}.

\bibitem{Wands:1998yp}
D.~Wands, ``{Duality invariance of cosmological perturbation spectra},'' Phys.\
  Rev.\ {\bf D60} (1999) 023507,
\href{http://arXiv.org/abs/gr-qc/9809062}{{\tt arXiv:gr-qc/9809062}}.

\bibitem{Finelli:2001sr}
F.~Finelli and R.~Brandenberger, ``{On the generation of a scale invariant
  spectrum of adiabatic fluctuations in cosmological models with a contracting
  phase},'' Phys.\ Rev.\ {\bf D65} (2002) 103522,
\href{http://arXiv.org/abs/hep-th/0112249}{{\tt arXiv:hep-th/0112249}}.

\bibitem{Brandenberger:2012zb}
R.~H. Brandenberger, ``{The Matter Bounce Alternative to Inflationary
  Cosmology},''
\href{http://arXiv.org/abs/1206.4196}{{\tt arXiv:1206.4196}}.

\bibitem{Xue:2010ux}
B.~Xue and P.~J. Steinhardt, ``{Unstable growth of curvature perturbation in
  non-singular bouncing cosmologies},'' Phys.\ Rev.\ Lett.\ {\bf 105} (2010)
  261301, \href{http://arXiv.org/abs/1007.2875}{{\tt arXiv:1007.2875}}.

\bibitem{Xue:2011nw}
B.~Xue and P.~J. Steinhardt, ``{Evolution of curvature and anisotropy near a
  nonsingular bounce},'' Phys.\ Rev.\ {\bf D84} (2011) 083520,
  \href{http://arXiv.org/abs/1106.1416}{{\tt arXiv:1106.1416}}.

\bibitem{Cai:2013vm}
Y.-F. Cai, R.~Brandenberger, and P.~Peter, ``{Anisotropy in a Nonsingular
  Bounce},'' Class.\ Quant.\ Grav.\ {\bf 30} (2013) 075019,
  \href{http://arXiv.org/abs/1301.4703}{{\tt arXiv:1301.4703}}.

\bibitem{Cai:2008qw}
Y.-F. Cai, T.-T. Qiu, R.~Brandenberger, and X.-M. Zhang, ``{A Nonsingular
  Cosmology with a Scale-Invariant Spectrum of Cosmological Perturbations from
  Lee-Wick Theory},'' Phys.\ Rev.\ {\bf D80} (2009) 023511,
\href{http://arXiv.org/abs/0810.4677}{{\tt arXiv:0810.4677}}.

\bibitem{Lin:2010pf}
C.~Lin, R.~H. Brandenberger, and L.~Perreault Levasseur, ``{A Matter Bounce By
  Means of Ghost Condensation},'' JCAP {\bf 1104} (2011) 019,
\href{http://arXiv.org/abs/1007.2654}{{\tt arXiv:1007.2654}}.

\bibitem{Brandenberger:2009yt}
R.~Brandenberger, ``{Matter Bounce in Horava-Lifshitz Cosmology},'' Phys.\
  Rev.\ {\bf D80} (2009) 043516,
\href{http://arXiv.org/abs/0904.2835}{{\tt arXiv:0904.2835}}.

\bibitem{Biswas:2005qr}
T.~Biswas, A.~Mazumdar, and W.~Siegel, ``{Bouncing universes in string-inspired
  gravity},'' JCAP {\bf 0603} (2006) 009,
\href{http://arXiv.org/abs/hep-th/0508194}{{\tt arXiv:hep-th/0508194}}.

\bibitem{WilsonEwing:2012pu}
E.~Wilson-Ewing, ``{The Matter Bounce Scenario in Loop Quantum Cosmology},''
  JCAP {\bf 1303} (2013) 026,
\href{http://arXiv.org/abs/1211.6269}{{\tt arXiv:1211.6269}}.

\bibitem{Brandenberger:2013zea}
R.~H. Brandenberger, C.~Kounnas, H.~Partouche, S.~P. Patil, and N.~Toumbas,
  ``{Cosmological Perturbations Across an S-brane},'' JCAP {\bf 1403} (2014)
  015,
\href{http://arXiv.org/abs/1312.2524}{{\tt arXiv:1312.2524}}.

\bibitem{Peter:2006hx}
P.~Peter, E.~J. Pinho, and N.~Pinto-Neto, ``{A Non inflationary model with
  scale invariant cosmological perturbations},'' Phys.\ Rev.\ {\bf D75} (2007)
  023516,
\href{http://arXiv.org/abs/hep-th/0610205}{{\tt arXiv:hep-th/0610205}}.

\bibitem{Odintsov:2014gea}
S.~Odintsov and V.~Oikonomou, ``{Matter Bounce Loop Quantum Cosmology from
  $F(R)$ Gravity},''
\href{http://arXiv.org/abs/1410.8183}{{\tt arXiv:1410.8183}}.

\bibitem{Cai:2014zga}
Y.-F. Cai and E.~Wilson-Ewing, ``{Non-singular bounce scenarios in loop quantum
  cosmology and the effective field description},'' JCAP {\bf 1403} (2014) 026,
\href{http://arXiv.org/abs/1402.3009}{{\tt arXiv:1402.3009}}.

\bibitem{Bamba:2014mya}
K.~Bamba, A.~N. Makarenko, A.~N. Myagky, and S.~D. Odintsov, ``{Bouncing
  cosmology in modified Gauss-Bonnet gravity},'' Phys.\ Lett.\ {\bf B732}
  (2014) 349--355,
\href{http://arXiv.org/abs/1403.3242}{{\tt arXiv:1403.3242}}.

\bibitem{Cai:2014jla}
Y.-F. Cai and E.~Wilson-Ewing, ``{A $\Lambda$CDM bounce scenario},'' JCAP {\bf
  1503} (2015) 006,
\href{http://arXiv.org/abs/1412.2914}{{\tt arXiv:1412.2914}}.

\bibitem{deHaro:2014kxa}
J.~de~Haro and J.~Amoros, ``{Viability of the matter bounce scenario in Loop
  Quantum Cosmology from BICEP2 last data},'' JCAP {\bf 1408} (2014) 025,
\href{http://arXiv.org/abs/1403.6396}{{\tt arXiv:1403.6396}}.

\bibitem{Elizalde:2014uba}
E.~Elizalde, J.~Haro, and S.~D. Odintsov, ``{Quasimatter domination parameters
  in bouncing cosmologies},'' Phys.\ Rev.\ {\bf D91} (2015) 063522,
\href{http://arXiv.org/abs/1411.3475}{{\tt arXiv:1411.3475}}.

\bibitem{deHaro:2015wda}
J.~de~Haro and Y.-F. Cai, ``{An Extended Matter Bounce Scenario: current status
  and challenges},''
\href{http://arXiv.org/abs/1502.03230}{{\tt arXiv:1502.03230}}.

\bibitem{Heard:2002dr}
I.~P. Heard and D.~Wands, ``{Cosmology with positive and negative exponential
  potentials},'' Class.\ Quant.\ Grav.\ {\bf 19} (2002) 5435--5448,
\href{http://arXiv.org/abs/gr-qc/0206085}{{\tt arXiv:gr-qc/0206085}}.

\end{thebibliography}
\end{document}